\documentclass[11pt,a4paper]{article}
\usepackage[centertags]{amsmath}
\usepackage{amsfonts}
\usepackage{amsthm}
\usepackage{amssymb,latexsym,a4wide}
\usepackage[T2A]{fontenc}
\usepackage[cp1251]{inputenc}
\usepackage[russian]{babel}
\usepackage[dvips]{graphicx}

\newtheorem{formula}{}[section]
\newtheorem{proposition}[formula]{Proposition}

\begin{document}

\baselineskip19pt

\begin {center}

{\bf Quantum optical device...}

       {\bf   Quantum optical device accelerating dynamic programming

 }
\vspace{3cm}

        {\bf D. Grigoriev $^1$, \quad A. Kazakov $^2$, \quad S. Vakulenko $^3$ }
\vspace{0.5cm}

$^1$IRMAR Universit\'e de Rennes, Beaulieu, 35042, Rennes, France

$^2$ Institute for Aerospace Instrumentation, St. Petersburg, Russia

        $^3$ Institute of Mechanical Engineering
         Problems, S. Petersburg, Russia
\vspace{0.2cm}

\vspace{3cm}

Address for correspondence: Grigoriev D., IRMAR Universit\'e de
Rennes, Beaulieu, 35042, Rennes, France

\vspace{5cm}

\end{center}

\begin{abstract}

In this paper  we discuss analogue computers based on quantum
optical systems accelerating dynamic programming for some
computational problems.
 These computers, at least in principle, can be realized
by actually existing devices.
 We estimate  an acceleration in resolving of some NP-hard
 problems that can be obtained in such a way
versus  deterministic computers.
\end{abstract}

\vspace{0.5cm}


\section{Introduction}
Quantum computing is intensive developing intersection of physics
and informatics since (Feynman 1982; Shor 1994; Grover  1997;
Preskill  1998). But schemes of universal quantum computers
working by qubits meet formidable difficulties in their
realization as real physical devices (Dyakonov 2003). In this
note we develop another approach based on analogue realization of
quantum computing. We concentrate our attention on some
particular NP-hard problems. For a given problem we can try to
construct a quantum analogue machine that resolves this problem
by means of time evolution.

Recall (Garey and Johnson  1979) that, roughly speaking, a problem
with instances $I$ lies in the class NP if there is a polynomial
time $P(|I|)$ algorithm checking  a solution (if this solution is
provided) where $|I|$ is the size of the input $I$. The problem
is NP-hard if $P=NP$ whenever this problem belongs to $P$. The
list of NP-complete problems contains many important problems of
number theory, graph theory, logic etc. Some of the most known
NP-complete problems are travelling salesman problem,
satisfiability, knapsack, matrix cover etc. (see a list in (Garey
and Johnson  1979)). Let us observe that all the NP-complete
problems, are, in a sense, polynomially equivalent. This means
that there are polynomial time reductions between different
problems.

An analogue quantum machine or resolving of
 combinatorial search NP-hard problems has  been proposed by (Farhi et al. 2000).
  Let us consider 3-SAT problem in which
  we deal with a family of $N$ clauses of the form
  $(B_{i_1} \ \bigvee \ B_{i_2} \ \bigvee \ \neg  \ B_{i_3})$
  or $(B_{i_1} \ \bigvee \ \neg \ B_{i_2} \ \bigvee \  \neg \ B_{i_3})$ or
  $(B_{i_1} \ \bigvee \ B_{i_2} \ \bigvee \ B_{i_3})$ etc.,
  where $B_{j}$ are boolean variables
  taking values True or False. The problem is whether one can assign such values
  to  all the variables that all the clauses will become true.

  The approach of (Farhi  et al. 2000) uses quantum adiabatic theorem.
 The authors describe a formal Hamiltonian  such that its ground
  state gives us  a solution of 3-SAT problem.
  The bound of the actual speed-up yielded by this algorithm depends on the
  spectral gap between the ground state and the first excited state. If this gap
  is exponentially small the time of solving is exponentially large.
  It is a very difficult problem to estimate the value of this gap
  and thus it is not easy to show, in general, that such an
  algorithm really gives a quantum acceleration.
  (Another NP-complete problem that can be naturally
  associated with a Hamiltonian
  is the problem of minimization of energy of spin glass with a large number of
  spins (see (Garey and Johnson 1979, p. 282.))

   The second difficulty with such approach is that it is not obvious to
realize physically  qubits  and Hamiltonian  with very nonlocal
nontrivial interaction between qubits.   It seems that
difficulties in physical realization of  Feynman quantum
computers or analogue computers from (Farhi et al. 2000)
 are significant and some authors even believe
(see, for example, (Dyakonov  2003)) that such computers are
physically non-realizable.

  The approach of (Farhi et al. 2000)
  uses a specific  structure of 3-SAT problem. If we use some universal approach
  that does not take into account a specific form of the problem,
  we can expect a quantum acceleration in time ${N}^{1/2}$ (Grover
1997), i.e., if a deterministic computer makes a search in $N$
steps then
 quantum universal computers will make the same search in $N^{1/2}$ steps. Further,
  it was shown that acceleration, with a universal approach can  be achieved by at
   most  $N^{1/6}$ {Beals et al. 1998; Preskill 1998}. Notice that this quantum
  $N^{1/2}$  acceleration is less than an acceleration that can be obtained by
  special deterministic and probabilistic  algorithms with
  respect to trivial   exhaustive search (Dantsin et al. 2003). Once more quantum machine
  based on quantum optics phenomena was discussed in (Kazakov 2003).
  The earlier analogue computational machines were discussed in
  (Matiyasevich 1987; Blass 1989).

In this paper we  describe a quantum machine
which, as we believe, can
be realized physically and
which may accelerate solving of some NP-complete problems.
Our machine uses different linear and nonlinear quantum optical devices
that can be constructed actually and, thus, our computer, at least in principle,
 can be practically realized.  An acceleration that can be obtained heavily depends
on different characteristics of our devices. We estimate this acceleration vs.
deterministic computers using the characteristics of actually existing devices.

\section{Statement of problems. Known results.}

We consider the following  problems which are NP-hard
(Papadimotriou, Steiglitz 1982; Garey, Johnson 1979).

{\bf 1} {\it Boolean knapsack, variant 1}

Given positive integers $c_j$, $j=1,...n$ and $K$, is there a
subset $S$ of $\{0,1,...,n\}$ such that $\sum_{j \in S} c_j =K$?

In other words, whether there exist  $n$ boolean values
$s_i \in \{0, 1 \}$ such that $\sum_{i=1}^n c_i s_i=K?$

Here the size $|I|$ of the  instance $\{n,c_1, c_2,...,c_n, K\}$
is the number of bits needed for binary
representations of all the integers $c_i, K$. We can suppose, without
loss of generality that $c_i < K$.
Thus,  size $|I|$ can be estimated roughly
 as $O(n\log K)$.

{\bf 2} {\it Boolean knapsack, variant 2}

Given integers $c_j$, $j=1,...n$ and $B_{-}, B_{+}$,
whether there exist  $n$ boolean values
$s_i \in \{0, 1\}$ such that $\sum_{i=1}^n c_i s_i$ lies in
the interval $(B_{-}, B_{+})$?
Here the instance size  is roughly $O(n\log B_{+})$.

{\bf 3} {\it Optimization boolean knapsack}

Given integers $c_j$ and $w_i$ , $j=1,...n$ and the number $B_{+}$,
 maximize the cost
\begin{equation}
 \sum_{i=1}^n w_i s_i
\label{twoone}
\end{equation}
defined by $n$ boolean variables $s_i \in \{0, 1 \}$ under
condition that
\begin{equation}
\sum_{i=1}^n c_i s_i < B_{+}.
\label{twotwo}
\end{equation}
There is an important difference
 between the problems {\bf 1, 2}  and {\bf 3}. The output in {\bf 1,2}
 is  "YES" or "NO". The output of {\bf 3} is a number, and one could try
 to approximate it.

Let us remind now some known results about  1, 2 and 3.

{\bf  a} All the  problems 1-3 can be resolved by exhaustive
search in $2^n$ steps.

{\bf b} If $2^n > K$ then the problem 1
 can be resolved by a more effective method,
namely, by {\it dynamic programming}, in $O(nK)$ steps. This
method can be described briefly as follows (for details, see
{Papadimotriou and Steiglitz 1982}). The algorithm produces
consecutively $\Sigma_0, \Sigma_1, ..., \Sigma_n$ such that
$\Sigma_j$ is the set of all possible subsums of $c_1,..., c_j$
 At the first step we set $\Sigma_0=\{0\}$.
At  $j+1$-th step, we set
$\Sigma_{j+1}=(\Sigma_j  \cup ( \Sigma_j +c_{j+1} )) \cap \{ 0,...,K \}$.
The problem has a solution if at the last $n$-th step
 $K \in \Sigma_n$.

In a similar way, we can resolve problem 2, it takes $O(nB_{+})$
steps, and problem 3, it takes $O(n \min(B_{+},R_{opt}))$ steps,
where $R_{opt}$ denotes the maximum cost (\ref{twoone}).

{\bf c} Suppose we solve an approximative problem 3, namely, we
seek for a number close to a maximal cost. This means that we
give up accuracy in exchange for  efficiency of our algorithm. To
this end we can apply a truncation procedure removing the last
$t$ digits from  the binary representations of $w_i$ and $c_i$.
Let  $w_m$ be the largest $w_i$. Such a procedure leads to a cost
$R_{appr}$ satifying the estimate (Papadimotriou and Steiglitz
1982; Ibarra and Kim 1975)
\begin{equation}
   \frac{R_{opt} -R_{appr}}{R_{opt}} < \epsilon, \quad \epsilon =\frac{n2^t}{w_m}
\label{twothree}
\end{equation}
The approximating  algorithm runs in time $O(n^4 \epsilon^{-1})$
(Papadimotriou and Steiglitz 1982; Ibarra and Kim 1975). Thus this
problem has an approximative solution that can be found in a
polynomial number of steps. Notice however, that there are many
NP-complete problems that cannot be approximated in such a way,
for example, travelling salesman problem (Garey and Johnson 1979;
Papadimotriou and Steiglitz 1982).

\section{Description of the quantum machines}

Consider $n+1$ points $x_0 < x_1 < x_2, ...< x_n = x_f$ located
along $x-$ axis in $(x,y)$ -plane. At the first point we set a
laser, which generates a narrow beam. The diameter of this beam
will be denoted by $d_b$, the wave length of the laser radiation
will be denoted by $\lambda$. Typical values of these parameters
are $d_b \sim 2\cdot 10^{-3} m, \ \lambda \sim 5 \cdot 10^{-7} m$.

Here we describe an analogue quantum optical device (QOD) for the
knapsack problems 1,2. Its possible scheme is presented on fig 1.
\begin{figure}[tp]
\begin{center}
\includegraphics [width=0.95\textwidth]{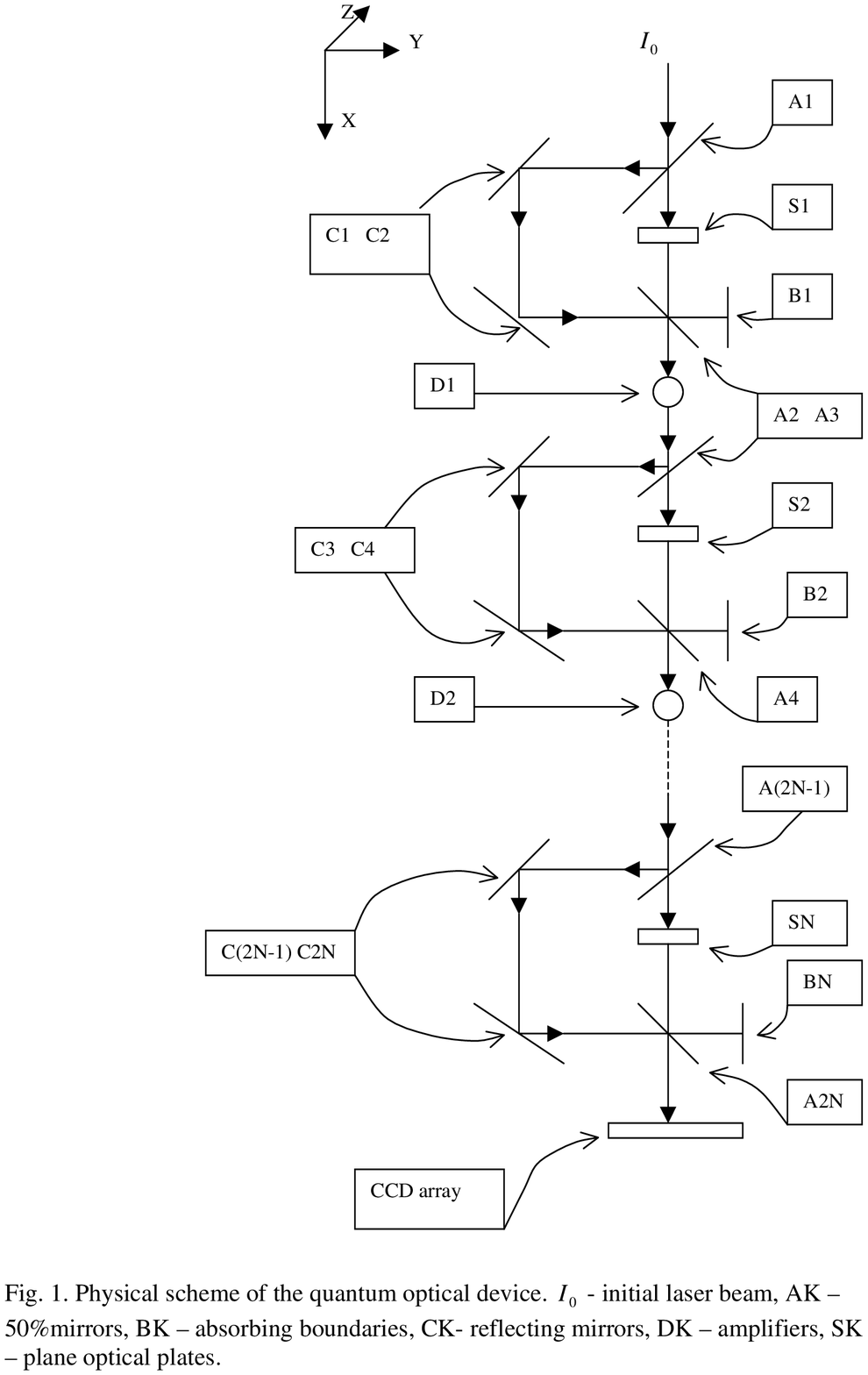}
\end{center}
\end{figure}

An input laser beam is splitted by $50\% $ mirror A1 to two
separate beams running two different trajectories. The beam 1 then
is shifted by a plane optical plate on the value $c_1 \kappa$ in
(vertical)  z-direction, which is perpendicular to the plane of
our picture (here $\kappa $ means the minimal shift). Then the
beams are on the second mirror B1 and united result (which is a
combination of shifted and non-shifted beams) goes through
amplifier C1. We suppose that this amplifier has the
characteristics presented on fig. 2.

\begin{figure}[tp]
\begin{center}
\includegraphics [width=0.9\textwidth]{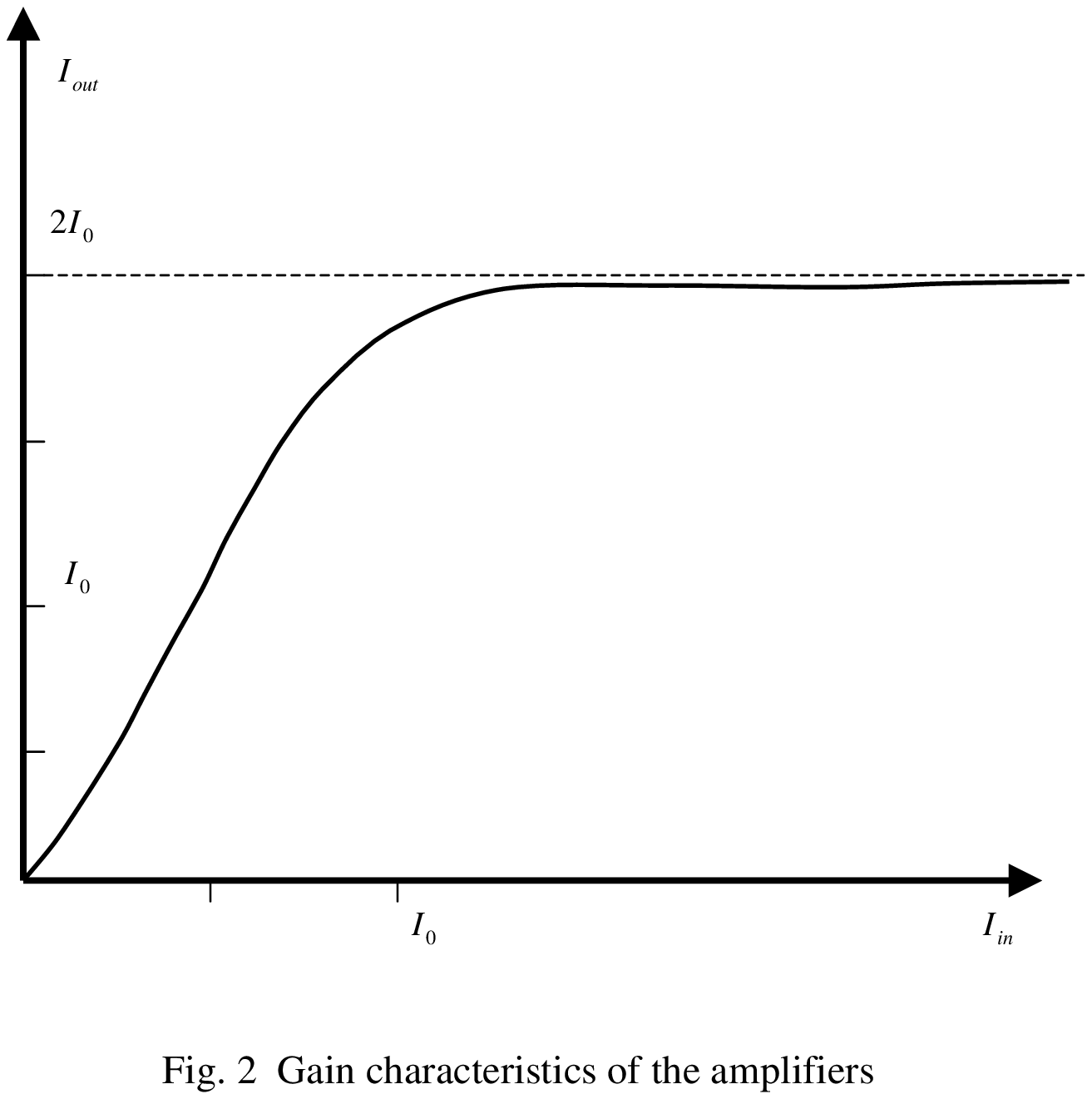}
\end{center}
\end{figure}

After passage of $m$ mirrors the propagating light will contain
beams shifted in  z-direction at all possible distances
$\sum_{i=1}^m c_i s_i \kappa$, where the values $s_i \in \{0,1 \}$
depend on the trajectory of the beam. At the final point we
measure intensity of outcoming light by a charge-coupled device
(CCD). A CCD camera uses a small rectangular piece
 of silicon  to receive incoming light. This silicon
 wafer is a solid state electronic component which has been
 micro-manufactured and segmented into an array of individual light-sensitive
 cells called "pixels". The pixel of  the most common CCD has size
 only about $\delta_p \approx 10^{-7} m\approx 0.2 \lambda$, it measures
 the intensity of light independently from other pixels.
Note, that namely $50\% $ mirrors are genuine quantum components
in our device (Feynman 1985).

 We denote by  $R_M$  the z-size of the separating mirrors
(and amplifiers). The last  parameter  important for estimating
of quality of our machine is an angular divergence $\alpha$ of
the laser beam. A usual laser produces a beam having the angular
divergence of order  $\alpha \sim d_b/\lambda \sim 4\cdot
10^{-5}$.

 Our machine is completely defined thus by the following parameters:
$$
 n,  R_M, L,  \delta, Y_{min}, Y_{max}, \lambda, d_b, \alpha,
 \kappa,
 c_i.
$$

For the problem 3, we use the following ideas. In order to
realize sums (\ref{twoone}, \ref{twotwo}) we have to modify
scheme presented on fig 1. Namely, together with z-shifts we use
the horizontal beam shift in an orthogonal direction $y$. Notice
that a beam shifted in $y$-direction contains this shift
propagating on different trajectories $L_1$ and  $L_2$ (see fig.
3). It means that after successive z- and y-shifts, which  the
initial beam gets propagating from $x_0$ to $x_n$, one obtains
the set of beams whose z- and y-shifts are different sums
$$
\sum_{i=1}^n c_i s_i, \quad \sum_{i=1}^n w_i s_i,
$$
in accordance with trajectories.

At the end of the device we set CCD which measure the plane
distribution of the outcoming light with z-coordinates $z < B_{+}
-\delta_p$, where $\delta_p$ is as it was mentioned above the
typical pixel size.

\begin{figure}[tp]
\begin{center}
\includegraphics [width=1.05\textwidth]{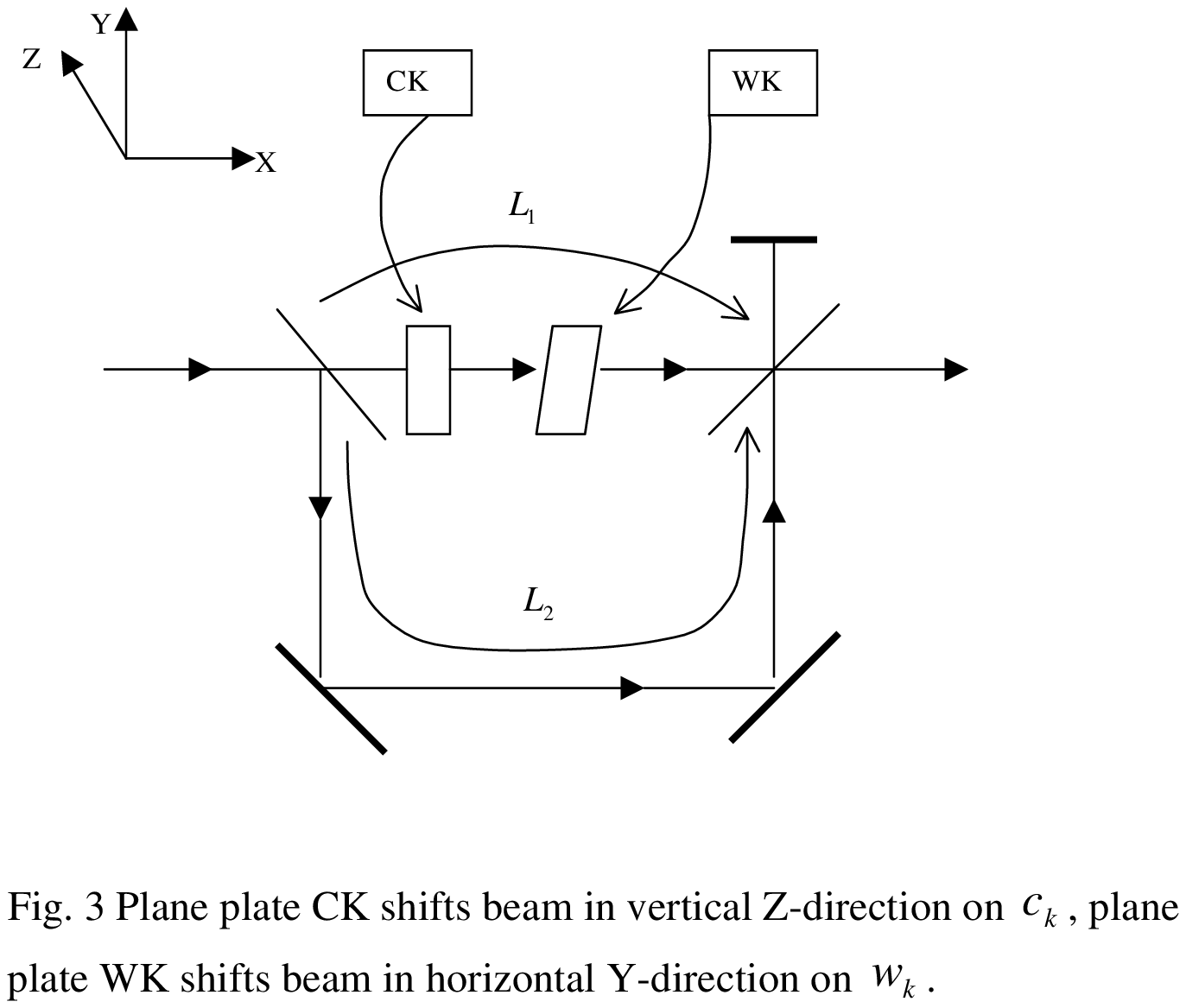}
\end{center}
\caption{}
\end{figure}

\section {Solving the dynamic programming
  by the quantum optical device}

Consider possible laser beam trajectories as a result of $n$
reflections on our mirrors. After the first series of reflections
we obtain two beams running along lines $y=h_1$ and $y=0$ since
each photon randomly chooses a way either along the first
trajectories or the second one. So, the mirror system serves as a
gate choosing the photon way. Remark that the mirror can be
considered simultaneously as a quantum and a classical device. In
a certain sense, our gates can be thus named as semiclassical
quantum gates.

 If we set a   CCD that registrates photons at  a point
 located after $x_1$, we could  registrate two localized
  z-separated gaussian beams at $y=0$ and $y=H_1$.

So, in an ideal situation, QOD works as follows. Consider the
problem 2. We set $H_i=\kappa c_i$, $i=1,2,...,n$, where $c_i, n$
are given. We take now $Y_{min}=\kappa B_{-}, \ Y_{max} =\kappa
B_{+}$. Now turning on  an input laser beam, we check, does the
interval $ (Y_{min}, Y_{max})$ contain the center of the any
laser beam (respectively, answer either "YES" or "

NO"). We
suppose that CCD solves the problem with an error less than $1/3$.

It is not difficult to see that this method of solution can be considered
as a physical realization of dynamic programming method
from Section 2.

The problem 3 can be resolved approximatively, i.e., the global
maximum can be estimated within some precision. However, to this
end, we must modify slightly our consideration (see below).

Consider now  diffraction effects. They can create an obstacle
since one could  registrate a photon  not really passing correct
gates as a photon that resolves our problem. In accordance with
theory of gaussian beams  (Svelto 1982)  each gaussian beam has a
finite transversal size and an angle of divergence $\alpha $.
These values are connected by
\begin{equation}
\alpha \sim \frac{1.2\lambda }{d_b}.\label{alpha}
\end{equation}
After $n$ reflections, the size of an output beam becomes
\begin{equation}
   d_{final}= n L \sin \alpha +d_b,
\label{final}
\end{equation}
where $L$ means distance between mirrors. Therefore, to get
minimal value of $d_{final} $ we set
\begin{equation}
d_b= \sqrt{nL\lambda }.
\label{diffr}
\end{equation}
Further, in order to separate on registering CCD the broadening
gaussian beams (whose amplitude can differ up to factor 2 in
accordance with fig.2), we have to restrict the minimal distance
$\kappa $ between beams by
\begin{equation}
\kappa \ge d_{final}=2d_b. \label{kappa}
\end{equation}

In addition to (4.3) we have a geometric restriction
\begin{equation}
     R_M > \kappa n + d_b.
\label{bound}
\end{equation}
 This inequality means that the shifts of no beams
jump out any mirrors.

One more problem is a possibility that, at  some step $i=T_0$ many
different beams arrive at the same point. It is not a
mathematical difficulty (since this means the existence of many
solutions) but it is a physical difficulty since it can lead to a
high energy concentration and destruction of the mirrors. To
avoid this effect we use active media that saturates the photon
density (see fig. 2). This allows us to restrict the photon
density at each point of the mirror  by some constant.

Using of the active media leads however to new difficulty which
can give us a lost of solutions. Namely, the phases of the
photons of each beam become slightly different after passing the
active media. Thus if the problem 2 has many solutions, CCD could
not registrate any photons as a result of interference. This
effect is possible if a phase shifts are significant. To overcome
this difficulty we use special auto phase control devices known
in optics. At the points $z_1, z_2,...z_n$ (see fig. 1) we places
the phase-adjusting devices which measure phase of all beams and
compare it  with phase of a reference laser beam of the same
frequency. Such adjustment gives the possibility to align phases
of mixing beams.

\section{Estimation of QOD performance for boolean knapsack }

Let us estimate now
 what we can do using such a QOD and compare this machine performance with
 deterministic computer performance.

\subsection{General estimates}

 There are  three important parameters: a) the implementation
 cost; b) the energy cost of solving; c) the running time
 when we solve $M$ times the same problem with different inputs.

 We compare here the dynamic programming
 for knapsack problems 1, 2 by a deterministic computer and by QOD. For the sake
 of unifying the notations we assume, that $B_+ \le K$.

 For the QOD the implementation cost $CI_{quant}$  can be estimated roughly as
 $$
 CI_{quant}=  O( K n)
 $$
 since the  mirror and amplifier
 sizes are of order $K$, the number of the mirrors and the cuvettes is $n$.
 The amplifier size is of order $Kn$, one can expect
 that energetical cost $CE_{qi}$ per one instance of the problem also is
 $$
 CE_{qi}= O( K n(n+K)),
 $$
 where factor $n+K$ arises due to greatest length of the photon trajectory.
 The complete energy involved in $M$ calculations is then
 $$
 CE_{quant}= O(K n(n+K) M).
 $$
 The implementation time has the order $ K n$ and the resolving time
 $$
     Time_{quant}\
le  C_1 (n+K)M,
 $$
 where $C_1$ is inverse proportional to the light speed and
thus this coefficient
 is rather small.
For the deterministic computer we have
 $$
 CE_{det} = O(KnM), \quad Time_{det} = O( KnM)
 $$
 Thus, QOD wins in time (taking into account both implementation
 and running times) vs. deterministic machine,
 while the order of consumed energy is higher than for
 a deterministic one.

 \subsection{Energetic and temporal costs}
 Our results include energetic and temporal costs of calculations.
 Let us consider a model, where this correspondence
 exhibits in a more explicit form. Let several polyhedra
  be placed in ${\bf R}^3$
 and we have to calculate the length of the shortest path
 from the point A to the point B
 avoiding these polyhedra (treated as obstacles)
 (see fig.4).  It is well known
 that this problem is NP-hard (Canny and Reif 1987).
 Moreover, even determining first $O(N^{1/2})$ bits of the length of the shortest
 path is NP-hard, where $N$ denotes the bits size of  description of the
 instance of the problem (Canny and Reif 1987).

\begin{figure}[tp]
\begin{center}
\includegraphics [width=1.03
\textwidth]{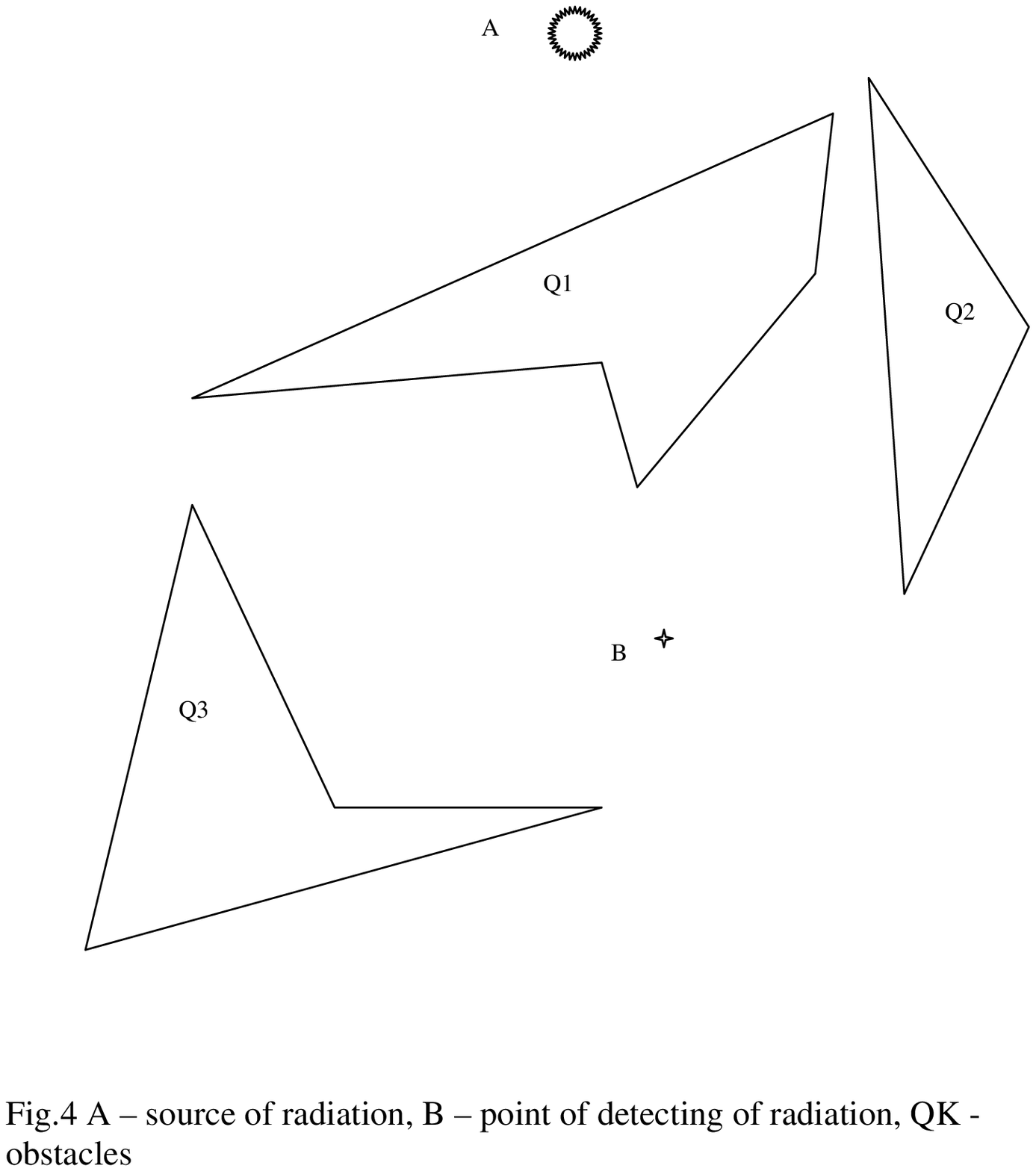}
\end{center}
\caption{}
\end{figure}

 We can realize this calculation by the
 following analogue physical device. Let us place in the point A
  a source of radiation, in the point B - detector and polyhedral obstacles
  have mirror boundaries.  If we turn the source A at initial
  moment $t=0$ and measure the time $t=t_1$ when our detector catches
  a first photon. Let the power of source be $J$ and frequency of
  radiation be $\nu $, then the number of created photons is
  $J/h\nu $, where $h$ is Planck's constant (Karlov 1992). Each photon moves
  in its path and   one can describe our analogue calculation as a "parallel
  computation" of the shortest path by different photons
  in which each photon plays the role of a processor.

  In general, this first photon caught by the detector
  could not belong to the front wave
  (which corresponds to the shortest paths)
  due to a weakening of the wave and possible presence of
the more strong waves.
  Therefore, increasing the intensity of the source allows to augment
  the probability to detect a photon of the wave front.
    So, we have to
  either repeat this experiment, with  corresponding increasing
 of temporal cost, or increase the intensity of the source with increasing
 of energetic cost of calculation.

 On the other hand, considered analogue devices have brought us to a conjecture that
  a tradeoff of the following kind
  $$
           E \cdot T \ge Z,
 $$
 holds,
 where $E$ and $T$ are energetical and temporal costs of
this calculation respectively,
and $Z$ depends on the problem.
This means (at some extent)
 that there could be an "exchange" between the energy cost  and
the temporal cost needed to solve a computational task.

\subsection{Estimates for real parameters}

 For the rate of a deterministic computer, for a single processor
 (we do not use a parallelism)
we take the value  $V=10^{10} operations/sec$. Suppose $2^n >
B_{+}$, then the  dynamic programming is more effective than the
exhaustive search. Then a deterministic processor solves the
problem 2 within time $T_c$
\begin{equation}
    T_c \approx n B_{+} V^{-1}
\label{fiveone}
\end{equation}
Consider our optical processor. We see that $B_{+}$ must be
subject to
\begin{equation}
       B_{+} < R_M/\kappa.
\label{fivetwoo}
\end{equation}
Thus our possibilities are restricted by the mirror size. Then
the problem can be resolved within time
\begin{equation}
   T_{q} \approx ( nL+R_M)/c + n T_{atom},
\label{fivethree}
\end{equation}
where the second term describes the time when photons spend in
the amplifier (which is at most $n$ times the relaxation time
$T_{atom}$ for active atoms, $T_{atom}$ is about $10^{-8} sec$).
 The preprocessing (implementation) time for the
optical machine is
$$
  T_{q, pre} \approx C_2 Kn,
$$

where $C_2$ is a constant.

 It is interesting now to see what we could obtain by using
 really existing devices. As an example, we take the following
 typical parameter values:
$$
R_M = nL=10 m, \quad d_b =2\cdot 10^{-3} m, \quad \lambda =5 \cdot
10^{-7} m.
$$

Then $\kappa=5 \cdot 10^{-3} m$ and $B_{+} < 2 \cdot 10^{2}$. For
$n \ge 30$ the next inequality holds, $2^n >> nB_+$, and dynamic
programming is more effective than the exhaustive search.

Minimal admissible difference $Y_{max}-Y_{min} $ can be estimated
as $\kappa=5 \cdot 10^{-3} m$. The deterministic time (5.1) is
then $ T_{det} \approx 30\cdot 2\cdot 10^{2} \cdot 10^{-10}
\approx 10^{-6} sec. $ In accordance with (\ref{fivethree}) $ T_q
\approx 30/(3 \cdot 10^{8}) + 30\cdot 10^{-8} \approx 10^{-6}
sec. $

So, our conclusions for problem 1, 2 are the following.
The performance of QOD is restricted by mirror sizes
and beam diameter. The problem 2 can be solved
by QOD and this effectiveness is essentially more
vs. a deterministic processor
 if we repeat the same
computation with different inputs many times.  For realistic
values of parameters, considered above, we can handle the case $n
< 60$, and then the speeds of the QOD and deterministic processor
have the same order.

Let us discuss now situation for problem 3. In the next
section  we describe first modifications needed for resolving
 this problem. We will see that in this case the speed-up  is much better.

\section{Estimate of machine performance for
optimization quantum knapsack }

\subsection{General estimates}
 We compare here the dynamic programming
 for knapsack problem 3 by a deterministic computer and by QOD
 using the parameters $CI$, $CE$, $Time$ described in Section 5.1.

 For the quantum optical device the implementation (preprocessing)
 cost $CI_{quant}$
 can be estimated
 roughly as
 $$
 CI_{quant} =O(K^2 n),
 $$
 since the the mirror size is now of order $K^2$, the number of the mirrors is $n$.
 The amplifier takes volume of order $K^2n$, one can expect
 that the energetical cost  $CE_{quant}$  is
 $$
 CE_{quant}=O( K^2 n(n+K) M),
 $$
 where now $K$ majorates sums (\ref{twoone}) and (\ref{twotwo}).
 However, if only numbers $w_i$ are large and $c_i \approx 1$, then
 the mirror area becomes again $O(K)$ as in Section 5, and
 we have the same estimates as in subsection 5.1.

 One can estimate the resolving time
 $$
     Time_{quant}=O( (n + K)M).
 $$

For the deterministic computer we can  (if we solve $M$ times the
problem with different inputs) estimate energy similarly to 5.1
(problems 1,2)
 $$
 CE_{det} =O( K nM), \quad Time_{det}=O( KnM).
 $$
 Thus, if the number of inputs $M >> K$
 then QOD has a chance to win, in time, with respect to a deterministic processor.
 The consumed energy of the QOD can be estimated by
 $$
  CE_{quant}=O(K^2n(n+K) M)
 $$
 (cf. subsection 5.1 above) whereas the energy consumed by a deterministic
 machine can be bounded again by $O( KnM)$ (cf. the discussion
 on the tradeoff of energy and time in subsection 5.1).

 Let us consider now approximating solutions (see Section 2).
 Recall that the processing time of the deterministic
 processor in order to obtain an approximative solution, within precision
$\epsilon$, is
$$
  Time_{det, appr} =O( M n^4 \epsilon^{-1}).
 $$
 To compare this performance  with effectiveness of our QOD, let us note
  that the pixel size is $\delta_p$ and thus the size of the mirrors
  in QOD solving the problem within relative
  precision $\epsilon$ should be
  $\delta_p/\epsilon= K\kappa $. Then the implementation cost will be
  $$
  CI_{quant} =O( (\delta _p/\epsilon \kappa )^2 n),
  $$
 resolving time can be estimated as $Time_{quant}=O(
(n+ \delta _p/\epsilon  \kappa )M)$
 The choice $\delta_p$ is restricted by a small number of possible values,
 thus, $Time_{quant} + CI_{quant}$ grows in $n$
 more slowly than $Time_{det, appr}$
for $\epsilon ^{-1} = o(n^3M) $.

The consumed energy of QOD for approximating problem
can be estimated by
$$
CE_{quant, appr} =O( (\delta _p/\epsilon  \kappa )^2 n(n + \delta
_p/\epsilon  \kappa )M)
$$
 which is less than the consumed energy for
a deterministic machine (which is of the order $n^4 \epsilon
^{-1}M$) when $\epsilon^{-1} = o((n \kappa /\delta _p)^{3/2})$.

\subsection{ Estimates for real device parameters}

Recall that, to resolve the problem 3, we have modified QOD
introducing additional parallel plates performing beam shifts in
the direction $y$. So, we can suppose that  $i$-th gate makes the
shift to $ H_i$ in the  direction $z$ and to
 $H^{\prime}_i$ to
the  $y$ axis. As above, we set $H_i = \kappa  c_i $ and
$H_i^{\prime} = \kappa  w_i $. At the end of the device we
situate a set of CCD, which measure the intensity of light in
$(y,z)$ plane.

 Let us describe now how we can resolve, approximatively, any problem
  3 without restrictions to the maximum of $B_{+}$.
  The key {\it truncation}
  idea can be found  in (Ibarra and Kim 1975), see also (Papadimotriou and Steiglitz 1982).
  We reduce the case of arbitrary coefficients
  $c_i < B_{+}, w_i$ to the previously studied
  case of restricted coefficients. To proceed it, we use the binary representations
  of these numbers. Suppose that each number use $\le m$ the digits $0,1$, i.e.
  the size of the binary input is not much than $m$.
  Moreover, assume that the maximal admissible value of
  $B_{+}$ can be written down with only $m_*$ digits. Now we truncate each given
  number $c_i, w_i$ and $B_{+}$ removing
  $m-m_*$ digits and allowing only first $m_*$ digits.
  With these new truncated data we can solve our problem and this solution
  gives an approximative solution.

The solving procedure to search an approximative solution of  the
problem 3 can be described as follows. We observe all beams that
are measured by CCD and between them we choose a beam that have a
maximal y-deviation. Such a scheme always gives an approximative
solution of our problem, with precision of order $\kappa /2$.

Let us compare now a deterministic processor and QOD. Let us take
the same parameters as above. The time processing for our device
will be chosen the same as above, i.e., $T_q= 10^{-6} sec$. For
the deterministic computer, according to subsection 2c, one has
$$
T_c \sim n^4 V\epsilon ^{-1},
$$
where $\epsilon$ is the relative precision. The relative error of
measurement we can estimate as $\epsilon \sim (\kappa /2)/R_M$,
which corresponds to the separation of the gaussian beams with
amplitudes differs by factor like 2. Substituting all the values
in the last relation, we have $ T_c \approx 1 sec $, that much
more than $T_q$.   So, the quantum optical  device gains an
acceleration $10^6$ times vs a typical deterministic computer.

\section{Conclusion}

We have introduced and designed analogous quantum optical devices
for simulating dynamic programming which provides the following
complexity bounds (see notations in subsections 5.1, 6.1) for
different versions of the knapsack problem (section 2).

\begin{proposition}
i) For the versions 1,2 the implementation cost
$
CI_{quant} = O(Kn)$, the running time $Time_{quant}= O((n+K) M)$,
and the energy cost $CE_{quant}=O(Kn(n+K)M)$;

ii) For the version 3
$
CI_{quant} = O(K^2n)$, the running time $Time_{quant}= O((n+K) M)$,
and the energy cost $CE_{quant}=O(K^2n(n+K)M)$;

iii) for $\epsilon$-approximative solution of the version 3
$
CI_{quant} = O(n(\delta _p/\epsilon )^2)$,
 the running time $Time_{quant}= O((n+\delta _p/\epsilon ) M)$,
and the energy cost $CE_{quant}=O(n(n+\delta /\epsilon )(\delta _p/\epsilon )^2 M)$;
(see subsections 5.1 and 6.1)
\end{proposition}

Also we compare these bounds with ones for deterministic machines (see
subsections 5.1, 6.1).

Let us discuss briefly the "quantum properties" of QOD. The
genuine quantum machine has to exploit two quantum properties: i)
"exponential resource" connected with  exponentially large
dimension of the state space, and ii ) the "quantum parallelism"
which is simultaneous evolution in all subspaces of the state
space.

The QOD described above contains only one genuine quantum
component, namely, $50 \%$ mirrors, which split the laser beams.
This splitting gives the possibility to realize the "quantum
parallelism". The splitting mirrors operate with the laser beams
that are macroscopic objects. However,  this macroscopity
prevents to the realization of "exponential resource".  In this
framework, realization of the exponential resource means using of
the mirrors of exponential size in order to separate the
exponentially large number of the laser beams. Moreover, in this
case we need exponentially large  energetic resource in order to
keep the intensity of the laser beams on a suitable level. So, it
is difficult to  realize of the "exponential resource with help
of macroscopic quantum devices. But the second quantum property -
"quantum parallelism" - can be realized by described above
devices.

\section{Ackowledgements}
The second and third authors are grateful to the Mathematical
Institute of  the University of Rennes, France, for the
hospitality.

\end{document}